\documentclass[12pt,a4paper]{article}
\usepackage{amssymb,amsmath}
\usepackage{amsfonts}
\usepackage{a4}
\numberwithin{equation}{section}
\begin{document}
	\baselineskip18pt
	
	\title{\bf Duadic negacyclic codes over a finite non-chain  ring and their Gray images}
	\author{Mokshi Goyal and Madhu Raka
		\\ \small{\em Centre for Advanced Study in Mathematics}\\
		\small{\em Panjab University, Chandigarh-160014, INDIA}\\
		\date{}}
	
\maketitle
{\abstract{Let $f(u)$ be a polynomial of degree $m, m \geq 2,$  which splits into distinct linear factors over a finite field $\mathbb{F}_{q}$. Let $\mathcal{R}=\mathbb{F}_{q}[u]/\langle f(u)\rangle$ be a finite non-chain ring. In an earlier paper, we studied duadic and triadic codes over $\mathcal{R}$ and their Gray images. Here, we study duadic negacyclic codes of Type I and Type II over the ring $\mathcal{R}$, their extensions and their Gray images. As a consequence some  self-dual, isodual,  self-orthogonal and complementary dual(LCD) codes over $\mathbb{F}_q$ are constructed. Some examples are also given to illustrate this.\vspace{2mm}
		}
	\section{ Introduction}
	
$~~~~~~$ Codes over finite rings have been known for several decades, but interest in these codes
increased substantially after a break-through work by Hammons et al. \cite{Ha} in 1994, which shows that some well known binary non-linear codes  can be constructed from  linear codes over $\mathbb{Z}_4$. Since then, a lot of research has been done on  cyclic codes, in particular on quadratic residue codes, over different types of finite rings, see for example \cite{CYY}, \cite{T}, \cite{Ka1}, \cite{Zh}, \cite{Ka2}, \cite{Li}, \cite{RKG}, \cite{GR1}.\vspace{2mm}

	Duadic cyclic codes or simply called duadic codes form an important class of cyclic codes. They generalize quadratic residue codes from prime to composite lengths.	In \cite{GR2}, the authors studied  duadic codes and their extensions over the ring $\mathbb{F}_{q}[u]/\langle u^m-u\rangle$, where $q$ is a prime power satisfying $q\equiv 1 ({\rm mod~}(m-1))$.  In another paper \cite{GR3}, the authors studied duadic and triadic codes over a more general non chain ring $\mathcal{R}=\mathbb{F}_{q}[u]/\langle f(u)\rangle$, where $f(u)$ is any polynomial of degree $m, m\geq 2$, which splits into distinct linear factors over $\mathbb{F}_{q}$. In that paper \cite{GR3}, the authors also defined a Gray map from $\mathcal{R}^n \rightarrow \mathbb{F}^{mn}_q$ which preserves duality and as a consequence obtained certain self-dual codes and isodual codes. \vspace{2mm}

 The concept of duadic cyclic codes was extended to  duadic negacyclic codes over $\mathbb{F}_{q}$ by Blackford \cite{Bl1} and then studied by Guenda \cite{Gu}. It has been further generalized to duadic constacyclic codes over $\mathbb{F}_{q}$ by Blackford \cite{Bl2} and by Fan and Zhang \cite{FZ}, where these authors have obtained necessary and sufficient conditions for existence of such codes. Chen et al. \cite{CDFL} studied the existence of  $m$-adic $\lambda$-constacyclic codes of Type I over $\mathbb{F}_{q}$.\vspace{2mm}

	 In this paper, we study duadic negacyclic codes of Type I and Type II  over the ring $\mathcal{R}$ in terms of their idempotent generators. The Gray images of these codes and  their extensions lead to construction of self-dual, isodual, self-orthogonal and complementary dual(LCD) codes over $\mathbb{F}_q$.  \vspace{2mm}
	
	The paper is organized as follows: In Section 2, we give some preliminaries, recall duadic negacyclic codes of length $n$ over $\mathbb{F}_q$   and study some of their properties.  In Section 3, we recall the ring $\mathcal{R} = \mathbb{F}_{q}[u]/\langle f(u)\rangle$ and  the Gray map $\Phi$ : $\mathcal{R}^n \rightarrow \mathbb{F}^{mn}_q$. In Section 4, we study duadic negacyclic  codes of Type I and Type II over $\mathcal{R}$ and their Gray images. We also give some examples to illustrate our results.
\section{Preliminaries}
	
	Let $R$ be a commutative ring with identity. An $R$-linear code $\mathcal{C}$ of length $n$ is an $R$- submodule of $R^n$. $\mathcal{C}$ is called $\lambda$-constacyclic for a unit $\lambda$ in $R$ if $\sigma_\lambda(\mathcal{C})\subseteq{\mathcal{C}}$, where\\\vspace{2mm} $~~~~~~~~~~~~~~~~~~~~~~~~~~~\sigma_\lambda((c_0,c_1,...,c_{n-1}))=(\lambda c_{n-1},c_0,c_1,...,c_{n-2})$\\
	If $\lambda=1$, $\mathcal{C}$ is cyclic. If $\lambda=-1$, $\mathcal{C}$ is called negacyclic. A constacyclic code $\mathcal{C}$ of length $n$ over $R$ can be regarded as an ideal of 	$R[x]/\langle x^n-\lambda\rangle$ under the correspondence \vspace{2mm}\\
	$~~~~~~~~~~~~~~c=(c_0,c_1,...,c_{n-1})\to c(x)=c_0+c_1x+...+c_{n-1}x^{n-1}({\rm mod ~}(x^n-\lambda))$. \vspace{2mm}\\
	If $R=\mathbb{F}_q$, a negacyclic code $\mathcal{C}$ of length $n$  has a unique generator polynomial $g(x)$ satisfying $g(x)|( x^n+1)$ and it has a unique idempotent generator $e(x)$ which is the multiplicative unity of the corresponding ideal in $\mathbb{F}_q[x]/\langle x^n+1\rangle$. If $R=\mathbb{F}_q$ and $n$ is odd, then negacyclic codes are scalar equivalent to cyclic codes under the map
	
	$$\mathbb{F}_q[x]/\langle x^n-1\rangle \to \mathbb{F}_q[x]/\langle x^n+1\rangle$$
$$a(x) \mapsto a(-x).$$
	Thus the theory of negacyclic codes of odd length is equivalent to the theory of cyclic codes of odd length. Therefore,  throughout this paper, we take $n$ to be \textbf{even}  and $q$  a power of an odd prime, coprime to $n$.\vspace{2mm}

\noindent	 Let $\delta$ be a primitive $2n$th root of unity in some extension field of $\mathbb{F}_q$. Then the roots of $x^n+1$ are $\delta^{2i+1}$ for $0\leq i\leq {n-1}$. Let $\mathcal{O}_{2n}$ be the set of odd integers from $1$ to $2n-1$. The defining set of negacyclic code $\mathcal{C}=\langle g(x) \rangle$ of length $n$ is the set $T= \{i \in \mathcal{O}_{2n} : \delta^i {\rm ~is~ a ~root ~of} ~ g(x)\}$. It is a union of $q$-cyclotomic cosets modulo $2n$. The dimension of $\mathcal{C}$ is $n-|T|$.  Aydin et al. \cite{Ay} gave a negacyclic BCH bound, i.e., if $T$ has $d-1$ consecutive odd integers, then the minimum distance is atleast $d$.\vspace{2mm}

\noindent For a linear code $\mathcal{C}$  over $\mathbb{F}_q$, the dual code  $\mathcal{C}^\bot$ is defined as $\mathcal{C}^\bot =\{x\in \mathbb{F}_q^n~ |~ x \cdot y=0 ~ {\rm for ~ all~} y \in \mathcal{C}\}$, where $x\cdot y$ denotes the usual Euclidean inner product. $\mathcal{C}$ is self-dual if $\mathcal{C}=\mathcal{C}^\bot$ and self-orthogonal if $\mathcal{C}\subseteq \mathcal{C}^\bot$. A code $\mathcal{C}$ is called isodual if it is equivalent to its dual $\mathcal{C}^\bot$. A linear code $\mathcal{C}$ whose dual $\mathcal{C}^\bot$ satisfies $\mathcal{C}\cap \mathcal{C}^\bot=\{0\}$ is called a complementary dual (LCD) code. \vspace{2mm}

\noindent Let $s$ be an integer  such that  $(s,2n)=1$. A multiplier $\mu_s$ is a map  from  $\mathcal{O}_{2n} \rightarrow \mathcal{O}_{2n}$ defined as $\mu_s(i)=si ({\rm mod~} 2n)$. It is  extended on $\mathbb{F}_q[x]/\langle x^n+1\rangle$ by defining  $\mu_s(a(x))=a(x^s) ({\rm mod~} x^n+1)$.\vspace{2mm}\\
 \noindent The following Lemmas are well known results; Lemma 1  is a  due to  \cite {Bl1}:\vspace{2mm}

		\noindent{\bf  Lemma 1}: \\ 1. If  $\mathcal{C}$ is a negacyclic code over $\mathbb{F}_q$ with defining set $T$ then $\mathcal{C}^{\perp}$, the dual of  $\mathcal{C}$, is again negacyclic with defining set $T^{\perp}= \{i \in \mathcal{O}_{2n} : -i({\rm mod }~2n)\not\in T\}= \mathcal{O}_{2n} -\mu_{-1}(T)$.\vspace{2mm}\\
2. If $n=2^an', a\geq 1$ for some odd integer $n'$, then self-dual negacyclic code over $\mathbb{F}_q$ of length $n$ exists if and only if $q \not\equiv -1 ({\rm mod}~2^{a+1})$. In particular,
 if $n$ is oddly even, i.e., if $a=1$ then self-dual negacyclic codes over $\mathbb{F}_q$ exists if and only if  $q\equiv 1({\rm mod}~4)$. \vspace{2mm}

%
%
%
	\noindent{\bf  Lemma 2}:
	
	\noindent(i)  Let $\mathcal{C}$ and $\mathcal{D}$ be negacyclic codes of length $n$ over a finite field $\mathbb{F}_q$ with defining sets $T_1$ and $T_2$ respectively. Then  $\mathcal{C}\cap \mathcal{D}$ and $\mathcal{C}+ \mathcal{D}$ are negacyclic codes with defining sets $T_1\cup T_2$ and $T_1\cap T_2$ respectively.\vspace{2mm}
	
	\noindent (ii) Let $\mathcal{C}$ and $\mathcal{D}$ be negacyclic codes of length $n$ over  $\mathbb{F}_q$ generated by the idempotents $E_1, E_2$ in $\mathbb{F}_q[x]/\langle x^n+1\rangle$, then $\mathcal{C}\cap \mathcal{D}$ and $\mathcal{C}+ \mathcal{D}$ are generated by the idempotents $E_1E_2$ and $E_1+E_2-E_1E_2$ respectively.\vspace{2mm}
	
	\noindent(iii) Let $\mathcal{C}$  be a negacyclic code of length $n$ over $\mathbb{F}_q$ generated by the idempotent $E$,  then  $\mu_a(\mathcal{C})$  is generated by $\mu_a(E)$ and $\mathcal{C}^{\perp}$ is generated by the idempotent $1-E(x^{-1})$.\vspace{2mm}
	
	\noindent \textbf{Remark:} Dual of a linear code over a finite ring  is defined in the same way and results in Lemma 2 (ii) and (iii) also hold true over any finite ring.
	
	 \subsection{ Duadic negacyclic Codes over $\mathbb{F}_q$}	
	
	 \noindent Suppose, we have $\mathcal{O}_{2n}=A\cup B \cup X  $, where \vspace{2mm}\\
	(i) $A$, $B$ and $X$ are union of $q$-cyclotomic cosets mod $2n$.\\
	(ii) $A, B$ and $X$ are pairwise disjoint.\\
	(iii) There exist a multiplier $ \mu_s$, $(s,2n)=1$ such that $\mu_{s}(A)= B$ and $\mu_{s}(B)= A$ and $\mu_{s}(X)= X$.\vspace{2mm}\\
	Then we say that a splitting of $n$ given by multiplier $\mu_{s}$ exists. If such a splitting exists, then codes having $A$, $B$, $A\cup X$ and $B\cup X$ as their defining sets are called negacyclic duadic codes.
	 A splitting is called of Type I if $X=\phi$ and is of Type II if $X=\{\frac{n}{2}, \frac{3n}{2}\}$.  Note that if  $q\equiv 1({\rm mod}~4)$, then $C_{\frac{n}{2}}= \{\frac{n}{2}\}$ and $C_{\frac{3n}{2}}= \{\frac{3n}{2}\}= C_{-\frac{n}{2}}$, but if  $q\equiv 3({\rm mod}~4)$, then $C_{\frac{n}{2}}= \{\frac{n}{2}, \frac{3n}{2}\}$. Here $C_a $ denotes the $q$-cyclotomic coset  $ \{ a,aq,aq^2,\cdots, aq^{m_a-1}\}$ where $m_a$ is the least positive integer satisfying $aq^{m_a}\equiv a({\rm mod}~2n)$. Further every multiplier $\mu_s$, $(s,2n)=1$ leaves the set $\{\frac{n}{2}, \frac{3n}{2}\}$ invariant. This is so because if $s\equiv 1({\rm mod}~4)$, $\mu_s(\frac{n}{2})= \frac{n}{2},$ and $ \mu_s(\frac{3n}{2})= \frac{3n}{2}$; whereas  if $s\equiv 3({\rm mod}~4)$, $\mu_s(\frac{n}{2})= \frac{3n}{2}$ and  $\mu_s(\frac{3n}{2})= \frac{n}{2}$.\vspace{2mm}

\noindent Blackford \cite{Bl1} showed that if $n$ is oddly even then there always exist a splitting of Type I or Type II.\vspace{2mm}
	
\subsubsection{\noindent {\bf Duadic negacyclic codes of Type I over $\mathbb{F}_q$}}\vspace{2mm}
	
\noindent 	If $\mu_{s}$ gives a splitting of Type I, then \vspace{2mm}

	$~~~~~~~~~~~~~~~~~~~~~~~~~~~~ x^n+1= A(x)B(x)$ \vspace{2mm}
	
	\noindent where $A(x)=\prod_{i\in A} (x-\delta^i), B(x)=\prod_{j\in B} (x-\delta^j)$ are polynomials over   $\mathbb{F}_q$.\vspace{2mm}
	
	\noindent $\mathbb{S}_1=\langle A(x)\rangle$ and $\mathbb{S}_2=\langle B(x)\rangle$ having $A$ and $B$ as defining sets are called duadic negacyclic codes of Type I. \vspace{2mm}
	
	\noindent{\bf Lemma 3}: Let $f_1$ and $f_2$ be idempotent generators of duadic negacyclic codes  $\mathbb{S}_1$ and $\mathbb{S}_2$ of Type I respectively. Then:\\	(i) $\mathbb{S}_1\cap\mathbb{S}_2=\{0\}$ and
	$\mathbb{S}_1+\mathbb{S}_2=\mathbb{F}_q[x]/\langle x^n+1\rangle$,\\
(ii) $f_1f_2=0, ~ f_1+f_2=1, \mu_s(f_1)=f_2, ~ \mu_s(f_2)=f_1.$\\
	(iii) If the splitting is given by $\mu_{-1}$, then $\mathbb{S}_1$ and $\mathbb{S}_2$ are self-dual codes.\\
	(iv) If $\mu_{-1}(\mathbb{S}_1)=\mathbb{S}_1$ and $\mu_{-1}(\mathbb{S}_2)=\mathbb{S}_2$, i.e., if the splitting is not given by $\mu_{-1}$, then $\mathbb{S}_1$ and $\mathbb{S}_2$ are isodual codes. \vspace{2mm}\\
	{\bf Proof:} By lemma 2 (i), the defining set of $\mathbb{S}_1\cap\mathbb{S}_2$ is $A\cup B$ and the defining set of $\mathbb{S}_1+\mathbb{S}_2$ is $A\cap B$. hence (i) follows. (ii) follows immediately from (i) using Lemma 2(ii)and (iii). Suppose splitting is given by $\mu_{-1}$. By Lemma 2 (iii), the idempotent generator of $\mathbb{S}^{\perp}_1$ is  $ 1 - \mu_{-1} (f_1(x))$ which is equal to $ 1- f_2(x)= f_1(x)$. Hence $\mathbb{S}^{\perp}_1= \mathbb{S}_1$. Similarly $\mathbb{S}^{\perp}_2= \mathbb{S}_2$. If splitting is not given by $\mu_{-1}$, then $ 1 - \mu_{-1} (f_1(x))= 1 -  f_1(x)= f_2(x)$.  Hence $\mathbb{S}^{\perp}_1=\mathbb{S}_2$. Similarly $\mathbb{S}^{\perp}_2=\mathbb{S}_1$.  Since $\mathbb{S}_1$ and $\mathbb{S}_2$ are equivalent codes $(\mu_{s}(\mathbb{S}_1)=\mathbb{S}_2)$, we find that $\mathbb{S}_1$ and $\mathbb{S}_2$ are isodual codes. $~~~~~~~~~~~~~~~~~~~~~~~~~~~~~\Box$\vspace{2mm}
	
	\noindent{\bf Remark:} If $q\equiv 1({\rm mod}~4)$, then Type I duadic negacyclic codes always exist with multiplier $\mu_{-1}$. This is so because for $1\leq a \leq 2n-1,$ $a$ odd, $a\not\equiv -aq^i ({\rm mod~}2n)$ for any $i$, as $n$ is even, so $C_a \neq C_{-a}$. Therefore if $C_a \subseteq A,$ we have $C_{-a} \subseteq \mu_{-1}(A)=B$. But if $q\equiv 3({\rm mod}~4)$, there exist no splitting of Type I as $C_{\frac{n}{2}}= \{\frac{n}{2}, \frac{3n}{2}\}$ is always contained in $X$. If $n$ is oddly even, a self-dual negacyclic code is duadic of Type I with multiplier $\mu_{-1}$ since for self-dual negacyclic codes to exist one must have $q\equiv 1({\rm mod}~(4))$.
	\vspace{2mm}
	
\subsubsection{\noindent {\bf Duadic negacyclic codes of Type II over $\mathbb{F}_q$}}\vspace{2mm}

\noindent If $\mu_{s}$ gives a splitting of Type II, then \vspace{2mm}
        	
        	$~~~~~~~~~~~~~~~~~~~~~~~~~~~~ x^n+1= A(x)B(x)(x^2+1)$ \vspace{2mm}
        	
        \noindent where $A(x)=\prod_{i\in A} (x-\delta^i), B(x)=\prod_{j\in B} (x-\delta^j)$ and $ x^2+1=(x-\delta^{\frac{n}{2}})(x-\delta^{\frac{3n}{2}})$ are polynomials over $ \mathbb{F}_q$. \vspace{2mm}

      \noindent    A polynomial $c(x)=  \sum_i c_ix^i\in \mathbb{F}_q[x]/\langle x^n+1\rangle$ is called even-like if $x^2+1$ divides it, i.e., if $c(\delta^{\frac{n}{2}})=0$ and  $c(\delta^{\frac{-n}{2}})=0$ which gives \vspace{2mm}

	  $~~~~c_0-c_2+c_4.....(-1)^\frac{n-2}{2}c_{n-2}=0$ and $c_1-c_3+c_5.....(-1)^\frac{n-2}{2}c_{n-1}=0$.\vspace{2mm}
	
	 \noindent  A negacyclic code $\mathbb{C}$ is called even-like if all its codewords are even-like otherwise it is called odd-like.
        	
%
	 \noindent
	The codes $\mathbb{D}_1=\langle A(x)\rangle $ and $\mathbb{D}_2=\langle B(x)\rangle$ having $A$ and $B$ as their defining sets are called a pair of odd-like duadic negacyclic codes and codes $\mathbb{C}_1=\langle A(x)(x^2+1)\rangle$ and $\mathbb{C}_2=\langle B(x)(x^2+1)\rangle$ having $A\cup X$ and $B\cup X$ as defining sets are called a pair of even-like duadic negacyclic codes. For $i=1,2$, let $e_i$  be idempotent generators of $\mathbb{C}_i$   and  $d_i$  be idempotent generators of $\mathbb{D}_i$.\vspace{2mm}

\noindent Let  ${p}(x)= \frac{x^n+1}{x^2+1}=(1-x^2+x^4-\cdots+x^{n-2})$ and $\overline{p}(x)=\frac{2}{n}(1-x^2+x^4-\cdots+x^{n-2}) $.\\ Note that $\overline{p}(x)$ is an idempotent in the ring
$\mathbb{F}_q[x]/\langle x^n+1\rangle$.  Further $\langle p(x)\rangle =\langle \overline{p}(x)\rangle $ is a negacyclic code over $ \mathbb{F}_q$ of dimension 2. \vspace{2mm}

\noindent Part of  Lemma 4  follows from   Theorem 11  of \cite{Bl1}. For the sake of completeness, we give its proof.\vspace{2mm}
	
	\noindent{\bf Lemma 4}: Let $\mathbb{C}_1$ and $\mathbb{C}_2$ be a pair of even-like negacyclic duadic codes of Type II over $\mathbb{F}_q$ with $\mathbb{D}_1$ and $\mathbb{D}_2$ the associated pair of odd-like negacyclic duadic codes. Then\\
		(i) $\mu_{s}(\mathbb{C}_1)=\mathbb{C}_2$, $\mu_{s}(\mathbb{C}_2)=\mathbb{C}_1$,\\
		(ii) $\mu_{s}(\mathbb{D}_1)=\mathbb{D}_2$, $\mu_{s}(\mathbb{D}_2)=\mathbb{D}_1$,\\
		(iii) $\mathbb{D}_1 + \mathbb{C}_2= \mathbb{D}_2 + \mathbb{C}_1 =\mathbb{F}_q[x]/\langle x^n+1\rangle$, \\(iv)  $\mathbb{D}_1 \cap \mathbb{C}_2= \mathbb{D}_2 \cap \mathbb{C}_1 =\{0\}$,\\ (v) $ \mathbb{D}_1 + \mathbb{D}_2 =\mathbb{F}_q[x]/\langle x^n+1\rangle$ and $\mathbb{D}_1 \cap \mathbb{D}_2= \langle p(x)\rangle$,\\ (vi) $\mathbb{C}_1\cap\mathbb{C}_2=\{0\}$ and $\mathbb{C}_1+\mathbb{C}_2=\langle x^2+1\rangle$,\\(vii) $d_1(x)=1-e_2(x)$, $d_2(x)=1-e_1(x)$, $d_1e_2=d_2e_1=e_1e_2=0,$ \\(viii) The idempotent generator of $\mathbb{D}_1 \cap \mathbb{D}_2$ is $\overline{p}(x)$, i.e., $d_1d_2= \overline{p}(x)$,\\ (ix) The idempotent generator  of $\mathbb{C}_1+\mathbb{C}_2$ is $1-\overline{p}(x)$, i.e., $e_1+e_2= 1-\overline{p}(x)$, \\
		(x) If $s=2n-1$, then $\mathbb{C}^{\perp}_1=\mathbb{D}_1$ and $\mathbb{C}^{\perp}_2=\mathbb{D}_2$, \\
		(xi) $\mathbb{D}_i=\mathbb{C}_i+\langle\overline{p}(x)\rangle=\langle\overline{p}(x)+e_i(x)\rangle$ for $i=1,2$ and  \\ (xii) $d_1+d_2= 1+\overline{p}(x)$, $d_1-e_1= \overline{p}(x)$, $d_2-e_2= \overline{p}(x)$. \vspace{2mm}\\
		{\bf Proof:} (i) and (ii) follows from the definition of duadic negacyclic codes of Type II. By lemma 2(i), we find that  the defining set of each of $\mathbb{D}_1 + \mathbb{C}_2$, $\mathbb{D}_2 + \mathbb{C}_1$  and  $\mathbb{D}_1 + \mathbb{D}_2$ is empty set. The defining set of  each of $\mathbb{D}_1 \cap \mathbb{C}_2$, $\mathbb{D}_2 \cap \mathbb{C}_1$   and  $\mathbb{C}_1 \cap \mathbb{C}_2$ is $\mathcal{O}_{2n}$. The defining set of $\mathbb{D}_1 \cap \mathbb{D}_2$ is $A\cup B$ and that of $\mathbb{C}_1 + \mathbb{C}_2$ is $X$. This gives (iii) to (vi). Then (vii) and (viii) follow from lemma 2(ii). To prove (ix), we note   the fact that $\mathbb{C}_1 + \mathbb{C}_2$ is the dual of $\mathbb{D}_1 \cap \mathbb{D}_2$; hence its idempotent generator is $1- \mu_{-1}(d_1d_2)= 1- \mu_{-1}(\overline{p}(x))= 1-\overline{p}(x)$.  If splitting is given by $\mu_{-1}$, then $\mathbb{C}^{\perp}_1=\langle 1-\mu_{-1}(e_1(x))\rangle=\langle 1-e_2(x)\rangle =\mathbb{D}_1 $ similarly $\mathbb{C}^{\perp}_2=\mathbb{D}_2$. This proves (x). As $\mathbb{C}_i$ and $\langle\overline{p}(x) \rangle $ are both subspaces of $\mathbb{D}_i$ and dimensions of codes on both sides is same, (xi)  follows from Lemma 2(ii) and the fact that $e_i(x)\overline{p}(x)=0$. The last statement follows from the previous ones. $~~~~~~~~~~~~~~~~~~~~~~~~~~~~~~~~~~~~~~~~~~~~~~~~~~~~~~~~~~~~~~~\Box$\vspace{2mm}

	\noindent{\bf Lemma 5:} Let $\mathbb{D}_1$ and $\mathbb{D}_2$ be a pair of odd-like duadic negacyclic codes of length $n$ over $\mathbb{F}_q$ with multiplier $\mu_{s}$ of Type II. Suppose $ 2+\gamma^2n=0$ has a solution $\gamma$ in $\mathbb{F}^{*}_q$. Let $\overline{\mathbb{D}}_i$ be the extension of $\mathbb{D}_i$,  for $i=1,2$, defined by \vspace{2mm} \\$\begin{array}{ll}\overline{\mathbb{D}}_i = \big \{ (c_0,c_1,\cdots,c_{n-1},c_{\infty},c_{\infty'}):  &c_{\infty}= \gamma\displaystyle \sum_{j=0}^{\frac{n-1}{2}}(-1)^{j} c_{2j},~~ c_{\infty'}=\gamma \displaystyle\sum_{j=0}^{\frac{n-1}{2}}(-1)^{j} c_{2j+1},\\&~~~~~~~~~~~~~~~~~~~~~~(c_0,c_1,\cdots,c_{n-1})\in \mathbb{D}_i \big \}. \vspace{2mm}\end{array}$\\
	Then the following hold:\\
	\noindent(i) If $s=2n-1$, then $\overline{\mathbb{D}}_i$ is self-dual for $i=1,2$,\\
	\noindent(ii) If $\mu_{-1}(\mathbb{D}_i)= \mathbb{D}_i$ for $i=1,2$, then $\overline{\mathbb{D}}_1^{\perp}=\overline{\mathbb{D}}_2$ and $\overline{\mathbb{D}}_2^{\perp}=\overline{\mathbb{D}}_1$. \vspace{2mm}\\	
This is Theorem 12 of \cite{Bl1}.	
	
		\section{The ring $\mathcal{R}$ and the Gray map}
		Let $q$ be a prime power, $q$=$p^{s}$. Throughout the paper, $\mathcal{R}$ denotes the commutative ring $\mathbb{F}_{q}[u]/\langle f(u)\rangle$, where $f(u)$ splits into distinct linear factors over $\mathbb{F}_{q}$. Let $f(u)=(u-{\alpha}_1)(u-{\alpha}_2)...(u-{\alpha}_m)$. $\mathcal{R}$ is a ring of size ${q}^m$ and characteristic $p$. Let $\eta_i ~;~ i= 1,2,\cdots,m $ denote the following elements of $\mathcal{R}$:
		\begin{equation} \begin{array}{ll}
		
			\eta_1= \frac{(u-\alpha_2)(u-\alpha_3)\cdots(u-\alpha_{m-1})(u-\alpha_m)}{(\alpha_1-\alpha_2)(\alpha_1-\alpha_3)\cdots(\alpha_1-\alpha_{m-1})(\alpha_1-\alpha_m)}\vspace{2mm}\\
			\eta_2= \frac{(u-\alpha_1)(u-\alpha_3)\cdots(u-\alpha_{m-1})(u-\alpha_m)}{(\alpha_2-\alpha_1)(\alpha_2-\alpha_3)\cdots(\alpha_2-\alpha_{m-1})(\alpha_2-\alpha_m)}\vspace{2mm}\\
			\cdots ~~~~~~~~~~~~~\cdots ~~~~~~~~~~~~\cdots ~~~~~~~~~~~~~~\cdots\vspace{2mm}\\
		\eta_i= \frac{(u-\alpha_1)(u-\alpha_2)\cdots(u-\alpha_{i-1})(u-\alpha_{i+1})\cdots(u-\alpha_m)}{(\alpha_i-\alpha_1)(\alpha_i-\alpha_2)\cdots(\alpha_i-\alpha_{i-1})(\alpha_i-\alpha_{i+1})\cdots(\alpha_i-\alpha_m)}\vspace{2mm}\\
		\cdots ~~~~~~~~~~~~~\cdots ~~~~~~~~~~~~\cdots ~~~~~~~~~~~~~~\cdots\vspace{2mm}\\
		\eta_m= \frac{(u-\alpha_1)(u-\alpha_2)\cdots(u-\alpha_{m-2})(u-\alpha_{m-1})}{(\alpha_m-\alpha_1)(\alpha_m-\alpha_2)\cdots(\alpha_m-\alpha_{m-2})(\alpha_m-\alpha_{m-1})}\vspace{2mm}\\
		
		\end{array}\end{equation}
		
		
		  \noindent One can easily find that
		  	\begin{equation} \begin{array}{ll}
		   $$\eta_i^2=\eta_i, ~\eta_i\eta_j=0~ {\rm ~for~} 1\leq i, j \leq m, ~i\neq j~{\rm~ and ~}\sum_{i=1}^m \eta_i=1$ in $\mathcal{R}$$.
		
		   	\end{array}\end{equation}
		\noindent The decomposition theorem of ring theory tells us that  $\mathcal{R}=\eta_1\mathcal{R}\oplus\eta_2\mathcal{R}\oplus \cdots \oplus\eta_m\mathcal{R}$.\vspace{2mm}
		
	\noindent Every element $r(u)$ of the ring $\mathcal{R}=\mathbb{F}_{q}[u]/\langle f(u)\rangle$ can be uniquely expressed as $$ r(u)=r_0+r_1u+r_2u^2+\cdots +r_{m-1}u^{m-1} = \eta_1a_1+\eta_2a_2+\cdots+\eta_m a_m$$ where $a_i = r(\alpha_i)$ for $i=1,2,...,m$. This is so because, by (1), $\eta_i(\alpha_i)=1$ and
	$\eta_i(\alpha_j)=0$ for all $j \neq i, 1\leq i, j \leq m$.\vspace{2mm}
	
	\noindent Let $\Phi$ be a Gray map defined by $\Phi : \mathcal{R}\rightarrow \mathbb{F}_q^m$   $$r(u)= \eta_1a_1+\eta_2a_2+\cdots+\eta_m a_m \longmapsto(a_1,a_2,\cdots, a_m)V$$
	where $V$ is any nonsingular matrix over $\mathbb{F}_q$ of order $m\times m$.
	This map can be extended from $\mathcal{R}^n$ to  $(\mathbb{F}_q^{m})^n$ component wise.\vspace{2mm}
	
	\noindent  For an element $r \in \mathcal{R}$, let the Gray weight be defined as $w_{G}(r) =w_H(\Phi(r))$, the Hamming weight of $\Phi(r)$. The  Gray weight of an element in $ \mathcal{R}^n$ and  Gray distance $d_{G}$ of  two elements in $ \mathcal{R}^n$ are defined in the natural way.\vspace{2mm}

\noindent The following theorem is a result of \cite{GR3}. For the sake of completeness we give a proof of it.\vspace{2mm}

		\noindent \textbf{Theorem 1.} The Gray map $\Phi$ is an  $\mathbb{F}_q$ - linear, one to one and onto map. It is also distance preserving map from ($\mathcal{R}^n$, Gray distance $d_{G}$) to ($\mathbb{F}_q^{mn}$, Hamming distance $d_H$). Further if the matrix $V$  satisfies $VV^T=\lambda I_m$, $\lambda \in \mathbb{F}_q^*$, where $V^T$ denotes the transpose of the matrix $V$, then the Gray image $\Phi(\mathcal{C})$ of a self-dual code $\mathcal{C}$  over $\mathcal{R}$ is a self-dual code in $\mathbb{F}_q^{mn}$. \vspace{2mm}\\		
			\noindent \textbf{Proof.} The first two assertions hold as $V$ is an invertible matrix over $\mathbb{F}_q$.\\ Let now $V=(v_{ij})$, $1\leq i,j\leq m$, satisfying $VV^T=\lambda I_m$. So that
		\begin{equation} {\displaystyle \sum_{k=1}^m}~ v_{jk}^2=\lambda ~~{\rm for ~all ~} j, 1\leq j \leq m {\rm ~~and ~~}  {\displaystyle \sum_{k=1}^m}~ v_{jk}v_{\ell k}=0 ~~{\rm for ~} j \neq \ell. \end{equation}
		Let $\mathcal{C}$ be a self-dual code over $\mathcal{R}$. Let $r=(r_0,r_1,\cdots,r_{n-1}), s=(s_0,s_1,\cdots,s_{n-1})$ $ \in \mathcal{C}$, where $r_i=\eta_1a_{i1}+\eta_2a_{i2}+\cdots+\eta_m a_{im}$ and $s_i=\eta_1b_{i1}+\eta_2b_{i2}+\cdots+\eta_m b_{im}$.  Using the properties of $\eta_i$'s from Lemma 5, we get
$$ r_i s_i = \eta_1a_{i1}b_{i1}+\eta_2a_{i2}b_{i2}+\cdots+\eta_m a_{im}b_{im}.$$ Then
$$0=r\cdot s=\sum_{i=0}^{n-1}r_is_i= \sum_{i=0}^{n-1}~\sum_{j=1}^{m}~\eta_j\hspace{0.5mm}a_{ij}\hspace{0.5mm}b_{ij}= \sum_{j=1}^{m}\eta_j \Big( \sum_{i=0}^{n-1}a_{ij}\hspace{0.5mm}b_{ij}\Big) $$  implies that
 \begin{equation}{\displaystyle \sum_{i=0}^{n-1}}a_{ij}b_{ij}=0, ~~~~~{\rm for ~ all ~} j=1,2,\cdots,m.\end{equation}
Now
		$$\Phi(r_i)=( a_{i1},a_{i2},\cdots,a_{im})V =\big(\sum_{j=1}^{m}~a_{ij}v_{j1},\sum_{j=1}^{m}~a_{ij}v_{j2}~,\cdots, \sum_{j=1}^{m}~a_{ij}v_{jm}\big)$$ Similarly $$\Phi(s_i) =\big(\sum_{\ell=1}^{m}~b_{i\ell}\hspace{0.5mm}v_{\ell 1},\sum_{\ell=1}^{m}~b_{i\ell}\hspace{0.5mm}v_{\ell 2}~,\cdots, \sum_{\ell=1}^{m}~b_{i\ell}\hspace{0.5mm}v_{\ell m}\big).$$  Using (3) and (4), we find that
$$\begin{array}{ll}\Phi(r)\cdot \Phi(s)&= {\displaystyle\sum_{i=0}^{n-1}}\Phi(r_i)\cdot \Phi(s_i)=  {\displaystyle\sum_{i=0}^{n-1}}~{\displaystyle\sum_{k=1}^m}~{\displaystyle\sum_{j=1}^{m}}~~{\displaystyle\sum_{\ell=1}^{m}}a_{ij}\hspace{0.5mm}
b_{i\ell}\hspace{0.7mm}v_{jk}\hspace{0.5mm}v_{\ell k}\vspace{2mm}\\&
		={\displaystyle\sum_{i=0}^{n-1}}{\displaystyle\sum_{j=1, \ell=j}^{m}}a_{ij}~b_{ij}\Big({\displaystyle\sum_{k=1}^m}~v_{jk}^2\Big)+{\displaystyle\sum_{i=0}^{n-1}}~{\displaystyle\sum_{j=1}^{m}}
		{\displaystyle\sum_{\ell=1,\ell\neq j}^{m}}a_{ij}\hspace{0.5mm}b_{i\ell}\Big({\displaystyle\sum_{k=1}^m}~v_{jk}\hspace{0.5mm}v_{\ell k}\Big)\vspace{2mm}\\&=\lambda {\displaystyle\sum_{i=0}^{n-1}}~{\displaystyle\sum_{j=1}^{m}}a_{ij}\hspace{0.5mm}b_{ij}\vspace{2mm}\\&=\lambda {\displaystyle\sum_{j=1}^{m}}\Big({\displaystyle\sum_{i=0}^{n-1}}~a_{ij}\hspace{0.5mm}b_{ij}\Big)=0,\end{array}$$
		 which proves the result. $~~~~~~~~~~~~~~~~~~~~~~~~~~~~~~~~~~~~~~~~~~~~~\Box$
	
		\section{Duadic negacyclic codes over the ring $\mathcal{R}$ }
		
			\noindent For a linear code $\mathcal{C }$ of length $n$ over the ring $\mathcal{R}$, let \vspace{2mm}\\
			$\mathcal{C }_1= \{ x_1\in \mathbb{F}_{q}^n : \exists ~x_2, x_3, \cdots, x_m \in \mathbb{F}_{q}^n {\rm ~such ~that~} \eta_1x_1+\eta_2x_2+\cdots+\eta_m x_m \in \mathcal{C }\},$\vspace{2mm}\\
			$\mathcal{C }_2= \{ x_2\in \mathbb{F}_{q}^n : \exists ~x_1, x_3, \cdots, x_m \in \mathbb{F}_{q}^n {\rm ~such ~that~} \eta_1x_1+\eta_2x_2+\cdots+\eta_mx_m \in \mathcal{C }\}$,\vspace{2mm}\\
			$~~~~~~~~\cdots ~~~~~~~~~~~~~\cdots~~~~~~~~~~~\cdots~~~~~~~~~~~~~~\cdots$\vspace{2mm}\\
			$\mathcal{C }_m= \{ x_m\in \mathbb{F}_{q}^n : \exists ~x_1,x_2, \cdots, x_{m-1} \in \mathbb{F}_{q}^n {\rm ~such ~that~} \eta_1x_1+\eta_2x_2+\cdots+\eta_mx_m \in \mathcal{C }\}$.\vspace{2mm}\\
			Then $\mathcal{C}_1,{C}_2,...,{C}_m$ are linear codes of length $n$ over $\mathbb{F}_{q}$,  $\mathcal{C}=\eta_1\mathcal{C}_1\oplus\eta_2\mathcal{C}_2\oplus\cdots \oplus\eta_m\mathcal{C}_m
			$ and $|\mathcal{C }|= |\mathcal{C }_1|~|\mathcal{C }_2|~\cdots~|\mathcal{C }_m|$.\vspace{2mm}
			
		\noindent	The following result is a simple generalization of a result of \cite{GR3}.\vspace{2mm}
			
			\noindent{\bf Theorem 2}: Let $\mathcal{C}=\eta_1\mathcal{C}_1\oplus\eta_2 \mathcal{C}_2\oplus \cdots\oplus \eta_m\mathcal{C}_m$ be a linear code of length $n$ over $\mathcal{R}$. Then \vspace{2mm}
			
			\noindent (i)~~ $\mathcal{C}$  is negacyclic over $\mathcal{R}$ if and only if $\mathcal{C}_i, ~i=1,2,\cdots,m$ are negacyclic over $\mathbb{F}_q$.\vspace{2mm}
			
			\noindent (ii) ~~If $\mathcal{C}_i=\langle  g_i (x)\rangle, ~g_i(x)\in \frac{\mathbb{F}_q[x]}{\langle x^{n}+1\rangle}$, $g_i(x)|(x^n+1)$,\\$~~~~~~$  then $\mathcal{C}=\langle \eta_1g_1(x),\eta_2g_2(x),\cdots,\eta_mg_m(x)\rangle =\langle g(x)\rangle\\ $~~~~~~~~~$~~~~$ where $g(x)= \eta_1g_1+\eta_2g_2+\cdots+\eta_mg_m$ and $g(x)|(x^{n}+1)$.\vspace{2mm}
			
			\noindent (iii)~~ Further $|\mathcal{C }|=q^{mn-\sum_{i=1}^{m}deg(g_i)}$.\vspace{2mm}
			
			\noindent (iv)~~ Suppose that $g_i(x)h_i(x)=x^n+1,~ 1\leq i\leq m.$ Let $ h(x)=\eta_1h_1(x)+\\~~~~~~~~~\eta_2h_2(x)+\cdots+\eta_mh_m(x),$ then
			$g(x)h(x)=x^n+1$. \vspace{2mm}
			
			\noindent(v)~~ $ \mathcal{C}^\perp=\eta_1\mathcal{C}_1^\perp\oplus\eta_2\mathcal{C}_2^\perp\oplus\cdots
			\oplus\eta_m\mathcal{C}_m^\perp.$ \vspace{2mm}
			
			\noindent (vi)~~ $ \mathcal{C}^\perp=\langle h^\perp(x)\rangle,$
			where $ h^\perp(x)=\eta_1h_1^\perp(x)+\eta_2h_2^\perp(x)+\cdots+\eta_mh_m^\perp(x)$,
			$~~~~~~~$where $h_i^\perp(x)$ is the reciprocal
			polynomial of $h_i(x), ~1\leq i\leq m.$  \vspace{2mm}
			
			\noindent(vii)$~~ |\mathcal{C}^\perp|=q^{\sum_{i=1}^m deg(g_i)}$.\vspace{2mm}
			
		\noindent We now define duadic negacyclic codes over the ring $\mathcal{R}$ in terms of their idempotent generators.
 Let $\mathcal{R}_n$ denote the ring $ \frac{\mathcal{R}[x]}{\langle x^{n}+1\rangle}$. Using the properties (2) of idempotents $\eta_i$, we have \\

		\noindent{\bf Lemma 6}: Let $\eta_i, ~1\leq i \leq m $ be idempotents as defined in (1). Then for  any tuple  $(E_{1}, E_{2}, \cdots, E_{m})$ of idempotents in the ring  $\frac{\mathbb{F}_q[x]}{\langle x^n+1\rangle}$,  $\eta_1E_{1}+\eta_2E_{2}+\cdots+\eta_m E_{m}$  is an idempotent in the ring $\mathcal{R}_n= \frac{\mathcal{R}[x]}{\langle x^{n}+1\rangle}$.
	\subsection{Duadic Negacyclic codes of Type I over the ring $\mathcal{R}$}
	Suppose there exists a splitting of $\mathcal{O}_{2n}$ of Type I over $\mathbb{F}_q$.  Let $f_1$ and $f_2$ be idempotent  generators of duadic negacyclic codes  $\mathbb{S}_1$ and $\mathbb{S}_2$ of Type I over $\mathbb{F}_q$. \vspace{2mm}

	\noindent Let the set $\{1,2,\cdots,m\}$ be denoted by  $\mathbb{A}$. For each $i \in \mathbb{A}$, let $F_{\{i\}}$ denote the idempotent of the ring $\mathcal{R}_n$ in which $f_1$ occurs at the $i$th place and $f_2$ occurs at the remaining $ 1,2,\cdots,i-1,i+1,\cdots,m$ places, i.e.
		\begin{equation}F_{\{i\}}=  \eta_1f_2+\eta_2f_2+\cdots+\eta_{i-1}f_2+\eta_if_1+\eta_{i+1}f_2+\cdots+\eta_mf_2= \eta_if_1+(1-\eta_i)f_2.\end{equation}
		
		
		\noindent In the same way, for $i_1,i_2,\cdots, i_k \in \mathbb{A}$,  $i_{r} \neq i_{s},~1\leq r,s \leq k$ let $F_{\{i_1,i_2,\cdots, i_k\}}$ denote the idempotent
		\begin{equation}F_{\{i_1,i_2,\cdots, i_k\}}=(\eta_{i_1}+\eta_{i_2}+\cdots +\eta_{i_k})f_1+(1-\eta_{i_1}-\eta_{i_2}-\cdots -\eta_{i_k})f_2.\end{equation}
		For $i \in \mathbb{A}$, $i_1,i_2,\cdots, i_k \in \mathbb{A}$, where $i_{r} \neq i_{s},~1\leq r,s \leq k$ let the corresponding idempotents be
		\begin{equation}F'_{\{i\}}=  \eta_if_2+(1-\eta_i)f_1.\end{equation}
		\begin{equation}F'_{\{i_1,i_2,\cdots, i_k\}}=(\eta_{i_1}+\eta_{i_2}+\cdots +\eta_{i_k})f_2+(1-\eta_{i_1}-\eta_{i_1}-\cdots -\eta_{i_k})f_1.\end{equation}
	
		\noindent Let $T_{\{i\}},~T'_{\{i\}},~ T_{\{i_1,i_2,\cdots, i_k\}}$ and $T'_{\{i_1,i_2,\cdots, i_k\}}$ denote the duadic negacyclic codes of Type I over $\mathcal{R}$ generated by the above defined  idempotents, i.e. \vspace{2mm}\\
		$T_{\{i\}}= \langle F_{\{i\}}\rangle $,~
		$T'_{\{i\}}= \langle F'_{\{i\}}\rangle $,~
		$T_{\{i_1,i_2,\cdots, i_k\}}= \langle F_{\{i_1,i_2,\cdots, i_k\}}\rangle $ and
		$T'_{\{i_1,i_2,\cdots, i_k\}}= \langle F'_{\{i_1,i_2,\cdots, i_k\}}\rangle $.\vspace{2mm}\\
		
		\noindent{\bf Theorem 3 :} For $i \in \mathbb{A}$, $T_{\{i\}}$ is equivalent to $ T'_{\{i\}}$. For $i_1,i_2,\cdots, i_k \in \mathbb{A}$,  $i_{r} \neq i_{s},~1\leq r,s \leq k$, $T_{\{i_1,i_2,\cdots, i_k\}}$ is equivalent to $ T'_{\{i_1,i_2,\cdots, i_k\}}$. Further
		there are $2^{m-1}-1$ inequivalent duadic negacyclic codes of Type I over the ring $\mathcal{R}$.\vspace{2mm}
		
		\noindent \textbf{Proof:}
		Let the multiplier $\mu_s$ give a splitting of $\mathbb{S}_1$ and $\mathbb{S}_2$. Then  $\mu_s(f_1)=f_2$, $\mu_s(f_2)=f_1$, so $\mu_s(\eta_if_1+(1-\eta_i)f_2) =\eta_if_2+(1-\eta_i)f_1$,  $\mu_s(F_{\{i_1,i_2,\cdots, i_k\}}) =F'_{\{i_1,i_2,\cdots, i_k\}}$, This proves that $T_i \sim T'_i$, $T_{\{i_1,i_2,\cdots, i_k\}}\sim T'_{\{i_1,i_2,\cdots, i_k\}}$. \vspace{2mm}
		
		\noindent Note that  $ F_{\mathbb{A}-\{i\}}= F'_{\{i\}}$, $ F_{\mathbb{A}-\{i_1,i_2,\cdots, i_k\}} = F'_{\{i_1,i_2,\cdots, i_k\}}$. Therefore
		\begin{equation} T_{\mathbb{A}-\{i\}} \sim T'_{\{i\}} \sim T_{\{i\}},  \vspace{-2mm}\end{equation} \begin{equation} T_{\mathbb{A}-\{i_1,i_2,\cdots, i_k\}}\sim T_{\{i_1,i_2,\cdots, i_k\}}. \end{equation}
		For a given positive integer $k$, the number of choices of the subsets $\{i_1,i_2,\cdots, i_k\}$ of $\mathbb{A}$ is $ m \choose k$.\vspace{2mm}
		
		Let $m$ be even first. Then $|\{i_1,i_2,\cdots, i_{m/2}\}|= |\mathbb{A}-\{i_1,i_2,\cdots, i_{m/2}\}|=\frac{m}{2}$. Using (9) and (10), we find that the number of inequivalent duadic negacyclic codes of Type I is $ {m \choose 1} + { m \choose 2}+\cdots { m \choose (m/2)-1}+ \frac{1}{2}{ m \choose m/2}=2^{m-1}-1$. If $m$ is odd the number of inequivalent duadic negacyclic codes of Type I is $ {m \choose 1} + { m \choose 2}+\cdots { m \choose (m-1)/2}=2^{m-1}-1$. $~~~~~~~~~~~~~~~~~~~~~~~~~~~~~~~~~~~~~~~~~~~~~~~~~~\Box$\vspace{4mm}

		Let $[x]$ denote the greatest integer $\leq x$. we have $[\frac{m}{2}]= \frac{m}{2}$, when $m$ is even and $[\frac{m}{2}]= \frac{m-1}{2}$, when $m$ is odd.\vspace{2mm}
		
		\noindent{\bf Theorem 4 :} For subsets ${\{i_1,i_2,\cdots, i_k\}}$ of $\mathbb{A} $ with cardinality $k$, $1 \leq k \leq [\frac{m}{2}] $,
		the following assertions hold for duadic negacyclic codes of Type I  over $\mathcal{R}$.
		\vspace{2mm}
		
		\noindent  (i)~~ $ T_{\{i_1,i_2,\cdots, i_k\}}\cap T'_{\{i_1,i_2,\cdots, i_k\}}= \{0\},$\vspace{2mm}\\
	\noindent	(ii)~~$ T_{\{i_1,i_2,\cdots, i_k\}}+T'_{\{i_1,i_2,\cdots, i_k\}}= \mathcal{R}_n,
	 $\vspace{2mm}\\
\noindent  (iii) If the splitting is given by $\mu_{-1}$, then $ T_{\{i_1,i_2,\cdots, i_k\}},~  T'_{\{i_1,i_2,\cdots, i_k\}}$  are  self-dual codes,\vspace{2mm}\\
        	 (iv) If the splitting is not given by $\mu_{-1}$, then $ T_{\{i_1,i_2,\cdots, i_k\}},~  T'_{\{i_1,i_2,\cdots, i_k\}}$
		 are isodual  codes.
		\vspace{2mm}
		
		\noindent \textbf{Proof:} From relations (2),(6) and (8), we see that $F_{\{i_1,i_2,\cdots, i_k\}}+F'_{\{i_1,i_2,\cdots, i_k\}}=f_1+f_2 $ and  $F_{\{i_1,i_2,\cdots, i_k\}}F'_{\{i_1,i_2,\cdots, i_k\}}=f_1f_2 $.  Therefore by Lemmas 2 and 3, \\$T_{\{i_1,i_2,\cdots, i_k\}}$ $\cap T'_{\{i_1,i_2,\cdots, i_k\}} $ $= \langle F_{\{i_1,i_2,\cdots, i_k\}}F'_{\{i_1,i_2,\cdots, i_k\}}\rangle = \{ 0\}$
	and  $T_{\{i_1,i_2,\cdots, i_k\}}+ T'_{\{i_1,i_2,\cdots, i_k\}}=  \langle F_{\{i_1,i_2,\cdots, i_k\}}+F'_{\{i_1,i_2,\cdots, i_k\}}-F_{\{i_1,i_2,\cdots, i_k\}}F'_{\{i_1,i_2,\cdots, i_k\}}\rangle$ $= \langle f_1+f_2-f_1f_2\rangle= \langle 1 \rangle = \mathcal{R}_n
	$. This proves (i) and (ii). \vspace{2mm}

If the splitting is given by $\mu_{-1}$, then by Lemma 3, $\mathbb{S}_1^{\perp} = \mathbb{S}_1 $ and $\mathbb{S}_2^{\perp} = \mathbb{S}_2 $. Therefore,  by Lemma 2, $1-\mu_{-1}(f_1(x))=f_1(x)$ and $1-\mu_{-1}(f_2(x))=f_2(x)$ i.e., $f_1(x^{-1})=1-f_1(x)$ and $f_2(x^{-1})=1-f_2(x)$. For $F_{\{i_1,i_2,\cdots, i_k\}}(x)=(\eta_{i_1}+\eta_{i_2}+\cdots +\eta_{i_k})f_1(x)+(1-\eta_{i_1}-\eta_{i_2}-\cdots -\eta_{i_k})f_2(x),$ $1- F_{\{i_1,i_2,\cdots, i_k\}}(x^{-1})=1-(\eta_{i_1}+\eta_{i_2}+\cdots +\eta_{i_k})(1-f_1(x))-(1-\eta_{i_1}-\eta_{i_2}-\cdots -\eta_{i_k})(1-f_2(x))= (\eta_{i_1}+\eta_{i_2}+\cdots +\eta_{i_k})f_1(x)+(1-\eta_{i_1}-\eta_{i_2}-\cdots -\eta_{i_k})f_2(x)=F_{\{i_1,i_2,\cdots, i_k\}}(x)$. This gives $ T_{\{i_1,i_2,\cdots, i_k\}}^{\perp} =  T_{\{i_1,i_2,\cdots, i_k\}}$. Similarly we have $ T'^{\perp}_{\{i_1,i_2,\cdots, i_k\}} =  T'_{\{i_1,i_2,\cdots, i_k\}}$. This proves (iii). \vspace{2mm}

If the splitting is not given by $\mu_{-1}$, then by Lemma 3, $\mathbb{S}_1^{\perp} = \mathbb{S}_2 $ and $\mathbb{S}_2^{\perp} = \mathbb{S}_1 $. Therefore,  by Lemma 2, $f_1(x^{-1})=1-f_2(x)$ and $f_2(x^{-1})=1-f_1 (x)$. Hence we get $1- F_{\{i_1,i_2,\cdots, i_k\}}(x^{-1})= (\eta_{i_1}+\eta_{i_2}+\cdots +\eta_{i_k})f_2(x)+(1-\eta_{i_1}-\eta_{i_2}-\cdots -\eta_{i_k})f_1(x)=F'_{\{i_1,i_2,\cdots, i_k\}}(x)$. This proves $T_{\{i_1,i_2,\cdots, i_k \}}^{\perp} =  T'_{\{i_1,i_2,\cdots, i_k \}}$ and hence (iv). $~~~~~~~~~~~~~~~~~~~~~~~~~~~~~~~~~~~~~~~\Box$
	
	\subsection{Duadic Negacyclic codes of Type II over the ring $\mathcal{R}$}

Suppose there exists a splitting of $\mathcal{O}_{2n}$ of Type II over $\mathbb{F}_q$.  Let $\mathbb{C}_1$ and $\mathbb{C}_2$ be a pair of even-like negacyclic duadic codes of Type II over $\mathbb{F}_q$ with $\mathbb{D}_1$ and $\mathbb{D}_2$ the associated pair of odd-like negacyclic duadic codes of Type II. For $i=1,2$, let $e_i$  be idempotent generators of $\mathbb{C}_i$   and  $d_i$  be idempotent generators of $\mathbb{D}_i$.\vspace{2mm}

	As in the previous section,  for $i_1,i_2,\cdots, i_k \in \mathbb{A}$,  $i_{r} \neq i_{s},~1\leq r,s \leq k$, let $D_{\{i_1,i_2,\cdots, i_k\}}$  and $D'_{\{i_1,i_2,\cdots, i_k\}}$ be  odd-like idempotents and $E_{\{i_1,i_2,\cdots, i_k\}}$  and $E'_{\{i_1,i_2,\cdots, i_k\}}$ be  even-like idempotents in the ring $\mathcal{R}_n= \frac{\mathcal{R}[x]}{\langle x^{n}+1\rangle}$ given by
	\begin{equation}D_{\{i_1,i_2,\cdots, i_k\}}=(\eta_{i_1}+\eta_{i_2}+\cdots +\eta_{i_k})d_1+(1-\eta_{i_1}-\eta_{i_2}-\cdots -\eta_{i_k})d_2,\end{equation}
	\begin{equation}D'_{\{i_1,i_2,\cdots, i_k\}}=(\eta_{i_1}+\eta_{i_2}+\cdots +\eta_{i_k})d_2+(1-\eta_{i_1}-\eta_{i_1}-\cdots -\eta_{i_k})d_1,\end{equation}
	\begin{equation}E_{\{i_1,i_2,\cdots, i_k\}}=(\eta_{i_1}+\eta_{i_2}+\cdots +\eta_{i_k})e_1+(1-\eta_{i_1}-\eta_{i_2}-\cdots -\eta_{i_k})e_2,\end{equation}
	\begin{equation}E'_{\{i_1,i_2,\cdots, i_k\}}=(\eta_{i_1}+\eta_{i_2}+\cdots +\eta_{i_k})e_2+(1-\eta_{i_1}-\eta_{i_2}-\cdots -\eta_{i_k})e_1.\end{equation}
		\noindent Let $ Q_{\{i_1,i_2,\cdots, i_k\}},~Q'_{\{i_1,i_2,\cdots, i_k\}}$ denote the odd-like duadic negacyclic codes of Type II and $ S_{\{i_1,i_2,\cdots, i_k\}},~S'_{\{i_1,i_2,\cdots, i_k\}}$ denote the even-like duadic negacyclic codes of Type II over $\mathcal{R}$ generated by the corresponding  idempotents, i.e. \vspace{2mm}
	
	$Q_{\{i_1,i_2,\cdots, i_k\}}= \langle D_{\{i_1,i_2,\cdots, i_k\}}\rangle $,~~
	$Q'_{\{i_1,i_2,\cdots, i_k\}}= \langle D'_{\{i_1,i_2,\cdots, i_k\}}\rangle $,\vspace{2mm}\\
	$~~~~~~S_{\{i_1,i_2,\cdots, i_k\}}= \langle E_{\{i_1,i_2,\cdots, i_k\}}\rangle $,~~
	$S'_{\{i_1,i_2,\cdots, i_k\}}= \langle E'_{\{i_1,i_2,\cdots, i_k\}}\rangle$.\vspace{2mm}

\noindent Working as in Theorem 3, we get \vspace{2mm}
	
	\noindent{\bf Theorem 5:}  For $i_1,i_2,\cdots, i_k \in \mathbb{A}$,  $i_{r} \neq i_{s},~1\leq r,s \leq k$, $Q_{\{i_1,i_2,\cdots, i_k\}}$ is equivalent to $ Q'_{\{i_1,i_2,\cdots, i_k\}}$ and  $S_{\{i_1,i_2,\cdots, i_k\}}$ is equivalent to $ S'_{\{i_1,i_2,\cdots, i_k\}}$. Further
	there are $2^{m-1}-1$ inequivalent odd-like duadic negacyclic codes of Type II and $2^{m-1}-1$ inequivalent even-like duadic negacyclic codes of Type II over the ring $\mathcal{R}$.\vspace{2mm}

	\noindent{\bf Theorem 6:} For subsets ${\{i_1,i_2,\cdots, i_k\}}$ of $\mathbb{A} $ with cardinality $k$, $1 \leq k \leq [\frac{m}{2}] $,
	the following assertions hold for duadic negacyclic codes over $\mathcal{R}$.
	\vspace{2mm}\\$\begin{array}{ll}
	
	{\rm (i)}& Q_{\{i_1,i_2,\cdots, i_k\}}\cap Q'_{\{i_1,i_2,\cdots, i_k\}}= \langle \overline{p}(x)\rangle,\vspace{2mm}\\
	{\rm (ii)}& Q_{\{i_1,i_2,\cdots, i_k\}}+Q'_{\{i_1,i_2,\cdots, i_k\}}= \mathcal{R}_n,  \vspace{2mm}\\
	{\rm (iii)}& S_{\{i_1,i_2,\cdots, i_k\}}\cap S'_{\{i_1,i_2,\cdots, i_k\}}= \{0\},\vspace{2mm}\\
	{\rm (iv)}& S_{\{i_1,i_2,\cdots, i_k\}}+ S'_{\{i_1,i_2,\cdots, i_k\}}= \langle 1-\overline{p}(x)\rangle, \vspace{2mm}\\
	{\rm (v)} &S_{\{i_1,i_2,\cdots, i_k\}}\cap \langle \overline{p}(x)\rangle = \{0\},~S'_{\{i_1,i_2,\cdots, i_k\}}\cap \langle \overline{p}(x)\rangle = \{0\},\vspace{2mm}\\
	{\rm (vi)} &S_{\{i_1,i_2,\cdots, i_k\}}+ \langle \overline{p}(x)\rangle = Q_{\{i_1,i_2,\cdots, i_k\}},~
	S'_{\{i_1,i_2,\cdots, i_k\}}+ \langle \overline{p}(x)\rangle = Q'_{\{i_1,i_2,\cdots, i_k\}},\vspace{2mm}\\
	{\rm (vii)}& |Q_{\{i_1,i_2,\cdots, i_k\}}|= q^{\frac{m(n+2)}{2}}, |S_{\{i_1,i_2,\cdots, i_k\}}|= q^{\frac{m(n-2)}{2}} \end{array}$\vspace{2mm}\\
where 	$\langle \overline{p}(x)\rangle$ and $\langle 1-\overline{p}(x)\rangle$ are regarded as negacyclic codes over the ring $\mathcal{R}$ of size $q^{2m}$.\vspace{2mm}\\
	\noindent \textbf{Proof:}
	From relations (2) and (11)-(14), we see that \\ $D_{\{i_1,i_2,\cdots, i_k\}}+D'_{\{i_1,i_2,\cdots, i_k\}}=d_1+d_2 $, $E_{\{i_1,i_2,\cdots, i_k\}}+E'_{\{i_1,i_2,\cdots, i_k\}}=e_1+e_2 $, \\ $D_{\{i_1,i_2,\cdots, i_k\}}D'_{\{i_1,i_2,\cdots, i_k\}}=d_1d_2 $ and $E_{\{i_1,i_2,\cdots, i_k\}}E'_{\{i_1,i_2,\cdots, i_k\}}=e_1e_2 $. \vspace{2mm}\\ Now using Lemmas 2 and 4, we get (i)-(iv). \vspace{2mm}

	\noindent Using that $\overline{p}(x)= (1-e_1-e_2)$ and $e_1e_2=0$ from  Lemma 4 and noting that $e_1^2=e_1, e_2^2=e_2$, we find that $E_{\{i_1,i_2,\cdots, i_k\}} \overline{p}(x)=0$.\vspace{2mm}\\
	Similarly using $e_1 +\overline{p}(x)=d_1$ and  $e_2 +\overline{p}(x)=d_2$ from  Lemma 4, we see that  $E_{\{i_1,i_2,\cdots, i_k\}}+ \overline{p}(x)=D_{\{i_1,i_2,\cdots, i_k\}}$.\vspace{2mm}\\ Therefore $S_{\{i_1,i_2,\cdots, i_k\}}\cap \langle \overline{p}(x)\rangle = \langle E_{\{i_1,i_2,\cdots, i_k\}} \overline{p}(x)
	\rangle= \{0\},$ and $S_{\{i_1,i_2,\cdots, i_k\}}+ \langle \overline{p}(x)\rangle  = \langle E_{\{i_1,i_2,\cdots, i_k\}} +\overline{p}(x)-E_{\{i_1,i_2,\cdots, i_k\}}\overline{p}(x)\rangle=
	\langle D_{\{i_1,i_2,\cdots, i_k\}}\rangle=Q_{\{i_1,i_2,\cdots, i_k\}}$.\vspace{2mm}\\ This proves (v) and (vi). \vspace{2mm}
	
	\noindent Finally for $1 \leq k \leq [\frac{m}{2}] $, we have $$|Q_{\{i_1,i_2,\cdots, i_k\}}\cap Q'_{\{i_1,i_2,\cdots, i_k\}}|= | \langle\overline{p}(x)\rangle|= q^{2m}.$$  Therefore
	$$ q^{mn}=|\mathcal{R}_n|=|Q_{\{i_1,i_2,\cdots, i_k\}}+ Q'_{\{i_1,i_2,\cdots, i_k\}}|=\frac{|Q_{\{i_1,i_2,\cdots, i_k\}}| |Q'_{\{i_1,i_2,\cdots, i_k\}}|}{|Q_{\{i_1,i_2,\cdots, i_k\}}\cap Q'_{\{i_1,i_2,\cdots, i_k\}}|} \vspace{-2mm}$$
$$~~~~~~~~~~~~~~~~~~~~~~~~~~~~~~~~~~~~~~~=\frac{|Q_{\{i_1,i_2,\cdots, i_k\}}|^2}{q^{2m}}.$$
	This gives $|Q_{\{i_1,i_2,\cdots, i_k\}}|=q^{\frac{m(n+2)}{2}}$.  Now   we find  that
\newpage
	$$ q^{\frac{m(n+2)}{2}}=|Q_{\{i_1,i_2,\cdots, i_k\}}|=|S_{\{i_1,i_2,\cdots, i_k\}}+ \langle\overline{p}(x)\rangle|=|S_{\{i_1,i_2,\cdots, i_k\}}| |\langle\overline{p}(x)\rangle|\vspace{0mm}$$
$$~~~~~~~~~~~~~~~~~~~~~~~~~~~~~~~~~~~~~~~~~~~~~~~~~=|S_{\{i_1,i_2,\cdots, i_k\}}| q^{2m}.$$
	since $ |S_{\{i_1,i_2,\cdots, i_k\}}\cap \langle\overline{p}(x)\rangle|= | \langle 0\rangle|= 1$. This gives $|S_{\{i_1,i_2,\cdots, i_k\}}|=q^{\frac{m(n-2)}{2}}$. \vspace{4mm}

	\noindent{\bf Theorem 7:} If  $\mu_{-1}(\mathbb{C}_1)=\mathbb{C}_2$, $\mu_{-1}(\mathbb{C}_2)=\mathbb{C}_1$, then
	for each possible tuple $\{i_1,i_2,\cdots, i_k\} \in \mathbb{A}$, the following assertions hold for duadic negacyclic codes over $\mathcal{R}$.
	\vspace{2mm}\\
	$\begin{array}{ll}
	{\rm (i)} & Q_{\{i_1,i_2,\cdots, i_k\}}^{\perp} =  S_{\{i_1,i_2,\cdots, i_k\}},  \vspace{2mm}\\
	{\rm (ii)}& S_{\{i_1,i_2,\cdots, i_k\}} {\rm ~is~ self-orthogonal}. \end{array} $\vspace{2mm}

	\noindent \textbf{Proof:} By    Lemmas 2 and  3, we have $\mathbb{C}_1$= $\mathbb{D}^{\perp}_1$=$\langle1-\mu_{-1}(d_1(x))\rangle$, so  $1-d_1(x^{-1})=e_1(x)$. Similarly $1-d_2(x^{-1})=e_2(x)$. Therefore we find that  $1- D_{\{i_1,i_2,\cdots, i_k\}}(x^{-1})=E_{\{i_1,i_2,\cdots, i_k\}}(x)$. Now result (i) follows from  Lemma 2. Using (vi) of Theorem 6, we have $S_{\{i_1,i_2,\cdots, i_k\}}\subseteq Q_{\{i_1,i_2,\cdots, i_k\}}= S_{\{i_1,i_2,\cdots, i_k\}}^{\perp}$. Therefore $S_{\{i_1,i_2,\cdots, i_k\}}$ is self-orthogonal.  ~~~~~~~~~~~~~~~~~~~~~~~~~~~~~~~~~~~~~~~~~~~~~~~~~~~~~~~~~~~~~$\Box$\vspace{2mm}
	
	\noindent Similarly we get\vspace{2mm}
	
	\noindent{\bf Theorem 8 :} If  $\mu_{-1}(\mathbb{C}_1)=\mathbb{C}_1$, $\mu_{-1}(\mathbb{C}_2)=\mathbb{C}_2$,    then
	for all possible choices of $\{i_1,i_2,\cdots, i_k\} \in \mathbb{A}$, the following assertions hold for duadic negacyclic codes over $\mathcal{R}$.\vspace{2mm}\\
	$\begin{array}{ll}{\rm (i)} &Q_{\{i_1,i_2,\cdots, i_k \}}^{\perp} =  S'_{\{i_1,i_2,\cdots, i_k \}}, \vspace{2mm}\\ {\rm (ii)}& Q'^{\perp}_{\{i_1,i_2,\cdots, i_k\}} =  S_{\{i_1,i_2,\cdots, i_k\}}. \end{array} $\vspace{2mm}\\

	The extended duadic negacyclic codes over $\mathcal{R}$ are formed in the same way as the extended duadic negacyclic codes over $\mathbb{F}_{q}$ are formed. \vspace{2mm}
	
	\noindent{\bf Theorem 9 :} Suppose there exists a $\gamma$ in $\mathbb{F}_q^* $ satisfying $~2+\gamma^{2}n=0$. If $\mu_{-1}(\mathbb{C}_1)=\mathbb{C}_2$, $\mu_{-1}(\mathbb{C}_2)=\mathbb{C}_1$, then
	for all possible choices of $\{i_1,i_2,\cdots, i_k\} \in \mathbb{A}$, the extended duadic negacyclic codes $\overline{Q_{\{i_1,i_2,\cdots, i_k\}}}$  of length $n+2$ are self-dual.\vspace{2mm}
	
	\noindent \textbf{Proof:} As $Q_{\{i_1,i_2,\cdots, i_k\}}=S_{\{i_1,i_2,\cdots, i_k\}}+\langle \overline{p}(x)\rangle$, by Theorem 6, let $\overline{Q_{\{i_1,i_2,\cdots, i_k\}}}$ be the extended duadic negacyclic code over $\mathcal{R}$ generated by

$$~~~~~~~\begin{array}{cccccccc}
		~~~~~~~~~~~~~ 0 & 1 & ~2 & ~~3&\cdots &  n-1 & \infty & \infty'
		\end{array}\vspace{-2mm}$$ $$ \overline{G_{\{i_1,i_2,\cdots, i_k\}}}=\left(
		\begin{array}{ccccccc}
		  & & &  &  &~ 0 &0\\
		  & & & G_{\{i_1,i_2,\cdots, i_k\}} & & ~0 &0 \vspace{2mm}\\
		~\vdots & & &  &  & ~ \vdots&\vdots \vspace{2mm}\\
		 ~1 & 0 &-1 &~0~~~\cdots ~~~~&  0 &\frac{n\gamma}{2}&0\\~0 & 1 &~0 &\hspace{-2mm}-1 ~~~\cdots ~~&  1&0&\frac{n\gamma}{2}
		\end{array}
		\right)\vspace{2mm}$$

\noindent	where $G_{\{i_1,i_2,\cdots, i_k\}}$ is a generator matrix for the even-like duadic negacyclic code $S_{\{i_1,i_2,\cdots, i_k\}}$. The row above the matrix shows the column labeling  by $\mathbb{Z}_n\cup \infty\cup \infty'$. Since the vector $v_1$ corresponding to ${p}(x)=(1-x^2+x^4-\cdots+x^{n-2})$ and $v_2$ corresponding to $x{p}(x)=(x-x^3+x^5-\cdots+x^{n-1})$ belong to $Q_{\{i_1,i_2,\cdots, i_k\}}$ and its dual $Q_{\{i_1,i_2,\cdots, i_k\}}^{\perp}$ is equal to $S_{\{i_1,i_2,\cdots, i_k\}}$, the last two rows of $\overline{G_{\{i_1,i_2,\cdots, i_k\}}}$ are orthogonal to all the previous rows of $\overline{G_{\{i_1,i_2,\cdots, i_k\}}}$. Also $(v_1,\frac{n\gamma}{2},0)\cdot (v_1,\frac{n\gamma}{2},0)=0$ and $(v_2,0,\frac{n\gamma}{2})\cdot (v_2,0,\frac{n\gamma}{2})=0$ as $\frac{n}{2}+\frac{\gamma^{2}n^{2}}{4}$ = 0 in $\mathbb{F}_q$. The last two rows are clearly orthogonal  as $(v_1,\frac{n\gamma}{2},0)\cdot (v_2,0,\frac{n\gamma}{2})=0$. Further as $S_{\{i_1,i_2,\cdots, i_k\}}$ is self-orthogonal by Theorem 8, we find that the code $\overline{Q_{\{i_1,i_2,\cdots, i_k\}}}$ is self-orthogonal. Now the result follows from the fact that $|\overline{Q_{\{i_1,i_2,\cdots, i_k\}}}|=q^{2m} |S_{\{i_1,i_2,\cdots, i_k\}}|=q^{\frac{m(n+2)}{2}}= |\overline{Q_{\{i_1,i_2,\cdots, i_k\}}}^{\perp}|$. ~~~~~~~~~~~~$\Box$\vspace{4mm}

\noindent Similarly we have \vspace{2mm}
	
	\noindent{\bf Theorem 10 :}  Suppose there exists a $\gamma$ in $\mathbb{F}_q^* $ satisfying $~2+\gamma^{2}n=0$. If  $\mu_{-1}(\mathbb{C}_1)=\mathbb{C}_1,~ \mu_{-1}(\mathbb{C}_2)=\mathbb{C}_2$, then  for all possible choices of $\{i_1,i_2,\cdots, i_k\} \in \mathbb{A}$, the extended duadic negacyclic  codes satisfy $\overline{Q_{\{i_1,i_2,\cdots, i_k\}}}^{\perp}=\overline{Q'_{\{i_1,i_2,\cdots, i_k\}}}$ and hence are isodual.\vspace{2mm}
	
	\noindent \textbf{Proof:}  Let ~$\overline{Q_{\{i_1,i_2,\cdots, i_k\}}}$~ and ~$\overline{Q'_{\{i_1,i_2,\cdots, i_k\}}}$~ be the extended duadic negacyclic codes over $\mathcal{R}$ generated by $G_1= \overline{G_{\{i_1,i_2,\cdots, i_k\}}}$ and by $G_2= \overline{G'_{\{i_1,i_2,\cdots, i_k\}}}$
%
%
%
%
 respectively, where $G_1$ is as defined in Thorem 9 and $G_2$ is same as $G_1$ except that it has  $ G'_{\{i_1,i_2,\cdots, i_k\}}$ in its top left corner in place of $ G_{\{i_1,i_2,\cdots, i_k\}}$.  Let $v_1$ and $v_2$ denote the vectors corresponding to polynomials $p(x)$ and $xp(x)$. As $v_1$ and $v_2$ belong to $Q'_{\{i_1,i_2,\cdots, i_k\}}$ and $Q'^{\perp}_{\{i_1,i_2,\cdots, i_k\}} =  S_{\{i_1,i_2,\cdots, i_k\}}$, $v_1$ and $v_2$ are orthogonal to all the rows of $ G_{\{i_1,i_2,\cdots, i_k\}}$. As in Theorem 9, the last two rows of $G_1$ are self-orthogonal and are orthogonal to each other. Further rows of $ G'_{\{i_1,i_2,\cdots, i_k\}}$ are in $S'_{\{i_1,i_2,\cdots, i_k\}}=Q^{\perp}_{\{i_1,i_2,\cdots, i_k\}}$, so are orthogonal to rows of $ G_{\{i_1,i_2,\cdots, i_k\}}$. Therefore all rows of $G_2$ are orthogonal to all the rows of $G_1$. Hence  $\overline{Q'_{\{i_1,i_2,\cdots, i_k\}}} \subseteq \overline{Q_{\{i_1,i_2,\cdots, i_k\}}}^{\perp}$.  Now the result follows from comparing their sizes.  \vspace{2mm}

	\noindent \textbf{Corollary :} Let the matrix $V$ taken in the definition of the Gray map $\Phi$ satisfy $VV^T=\lambda I_m$, $\lambda \in \mathbb{F}_q^*$. If $\mu_{-1}(\mathbb{C}_1)=\mathbb{C}_2$ , then  for all possible choices of $\{i_1,i_2,\cdots, i_k\} \in \mathbb{A}$, the Gray images of extended duadic negacyclic codes $\overline{Q_{\{i_1,i_2,\cdots, i_k\}}}$, i.e.,  $\Phi(\overline{Q_{\{i_1,i_2,\cdots, i_k\}}})$  are self-dual codes of length $m(n+2)$ over $\mathbb{F}_q$ and the Gray images of the even-like duadic negacyclic codes $S_{\{i_1,i_2,\cdots, i_k\}}$, i.e., $\Phi(S_{\{i_1,i_2,\cdots, i_k\}})$ are self-orthogonal codes of length $mn$ over $\mathbb{F}_q$. If $\mu_{-1}(\mathbb{C}_1)=\mathbb{C}_1$, then $\Phi(\overline{Q_{\{i_1,i_2,\cdots, i_k\}}})$  are isodual codes of length $m(n+2)$ over $\mathbb{F}_q$.\vspace{2mm}

In  Tables 1 and 2, we give some examples of Gray images of duadic negacyclic codes. The minimum distances of these codes have been computed using the software `MAGMA'. 		

\section{Conclusion}
In this paper,  duadic negacyclic codes of Type I and Type II and  their extensions  over a finite non-chain ring $\mathcal{R}=\mathbb{F}_{q}[u]/\langle f(u)\rangle$ are studied, where $f(u)$ is a polynomial of degree $m (\geq 2)$  which splits into distinct linear factors over $\mathbb{F}_{q}$. Their Gray images under the  Gray map : $\mathcal{R}^n$ to $(\mathbb{F}_{q}^m)^n$, which preserves  self duality of linear codes, lead to  self-dual, isodual,  self-orthogonal and complementary dual(LCD) codes over $\mathbb{F}_q$. Some examples are also given to illustrate this. Further  in this direction,  polyadic  constacyclic codes over the ring $\mathcal{R}$ can be explored.

\newpage
 \textbf{Table 1.} Gray images of Type I duadic negacyclic codes \vspace{6mm} \\
\footnotesize{	\begin{tabular}{|c|c|c|c|c|c|c|}
		\hline
		$q$ & $n$ & $m$ &$s$& $f(u)$ & $V$ &$\Phi({T}_{\{1\}})$ \\ \hline
		5 & 18 & 2 &-1& $(u-2)(u-4)$ & 2,3  & [36,18,4] \\
		&   &   & & &   -3,2 & self-dual \\ \hline
		5&    18 & 3 &-1& $(u-2)(u-3)(u-4)$& 4,4,-2 & [54,27,8]\\
		&        &   &&                           & -2,4,4  & self-dual\\
		&        &   & &                          & 4,-2,4  & \\ \hline
		5 &   22  & 4 & -1 & $u(u-1)$ & 1,12,1,1  & [88,44,12]\\
		&     &      && $(u-3)(u-4)$ & -12,1,1,-1 & self-dual\\
		&      &     &&                                                    & -1,-1,1,12  &\\
		&      &     &&                                                    & -1,1,-12,1  &\\ \hline
		$9^*$ & 4 & 2 &3& $(u-\alpha)(u-\alpha^2)$ & 1,$\alpha$  & [8,4,4] \\
		&   &   & & &   $-\alpha,1$ & isodual \\ \hline
			$9^*$ &   4  & 4   &3& $(u-\alpha)(u-\alpha^2)$& $\alpha,-\alpha^2,1,1$  & [16,8,6]\\
			&     &      &&$(u-\alpha^4)(u-\alpha^6)$ & $-1,1,\alpha,\alpha^2$ & isodual\\
			&      &     & &                                                   & $\alpha^2,\alpha,-1,1$  &\\
			&      &     & &                                                   & $1,1,\alpha^2,-\alpha$ &\\ \hline
		13 &  6 &  3 &-1& $(u+1)(u-3)(u-4)$ & 2,-2,1 & [18,9,6]\\
		&     &   & &                                       & 2,-2,1  & self-dual\\
		&     &   & &                                       & 2,1,-2  & \\ \hline
		13 &   6  & 4   &-1& $u(u-3)$ & 1,12,1,1  & [24,12,7]\\
		&     &      && $(u+3)(u+4)$ & -12,1,1,-1 & self-dual\\
		&      &     & &                                                   & -1,-1,1,12  &\\
		&      &     & &                                                   & -1,1,-12,1  &\\ \hline
		13 &   6  & 5   &-1& $u(u-3)$ &8,11,11,2,12  & [30,15,8]\\
		&     &      && $(u-4)(u-8)(u+2)$ & 12,8,11,11,2 & self-dual\\
		&      &     &&                                                    & 11,2,12,8,11  &\\
		&      &     & &                                                   & 11,2,12,8,11  &\\
		&      &     & &                                                   & 11,11,2,12,8  &\\ \hline

	\end{tabular}}\\
\noindent $^*$ Here $\alpha$ is a primitive element of $\mathbb{F}_9$.
	\newpage
	$~~~~~~$ \textbf{Table 2.} Gray images of Type II duadic negacyclic codes \vspace{6mm}\\
\footnotesize{\begin{tabular}{|c|c|c|c|c|c|c|c|c|}
		\hline
		$q$ & $n$ & $m$ &$s$&$\gamma$& $f(u)$ & $V$ & $\Phi(D_{\{1\}})$ &$\Phi(\overline{E}_{\{1\}})$ \\ \hline
		3 & 10 & 2 &-1&1& $u(u-2)$ & 1,2 & [20,8,6] & [24,12,6] \\
		&   &   &&&  &   -2,1 & self-orthogonal& self-dual \\\hline
		
		3 & 22 & 2 &-1&1& $(u-1)(u-2)$ & 1,2 & [44,20,9] & [48,24,9] \\
		&   &   &&&  &   -2,1 & self-orthogonal& self-dual \\ \hline
		3 & 22 & 3 &-1&1& $u(u-1)(u-2)$  & 1,0,0 & [66,30,6] & [72,36,6] \\
		&   &   &&&  &  0,1,0 & self-orthogonal& self-dual \\
		&   &   &&&  &  0,0,1&         & \\ \hline
		
		7 &  10  & 3 &-1&2& $(u-3)(u-4)$& 2,-2,1 & [30,12,8] &[36,18,6]\\
		&        &     &&&   $(u-6)$                                  & 1,2,2 & self-orhogonal & self-dual\\
		&        &      & &&                                    & 2,1,-2& & \\ \hline
		7 &   10  & 4   &-1&2& $(u-2)(u-3)$ & 2,-2,1,1 &[40,16,12] & [48,24,6]\\
		&     &      &&& $(u-4)(u-6)$ & 1,1,2,2& self-orthogonal & self-dual\\
		&      &     &&&                                                    & 2,2,1,-1 & &\\
		&      &     &&&                                                    & 1,1,-2,2 & &\\ \hline
		11 & 26 & 2 &-1&4& $(u-4)(u-5)$ & 1,1 & [54,24,10] & [56,28,10] \\
		&   &   &&&  &   -1,1 &self-orthogonal& self-dual \\\hline
		11 &   10  & 4   &-1&does & $(u-2)(u-7)$ & 1,3,1,9 &[40,16,8] & \\
		&     &      &&not& $(u-8)(u-10)$ & -3,1,9,-1& self-orthogonal & - \\
		&      &     && exist&                                                    & -1,-9,1,3 & &\\
		&      &     &&&                                                    & -9,1,-3,1 & &\\ \hline
			13 & 34 & 2 &9&4& $(u-1)(u-2)$ & 1,1 & [68,36,12] & [72,36,12] \\
			&   &   &&&  &   -2,1 & LCD& isodual \\\hline
				13 & 6 & 2 &5&2& $(u-2)(u-3)$ & 1,2 & [12,4,7] & [16,8,4] \\
				&   &   &&  & &  -2,1 & LCD& isodual \\\hline
		13 & 6 & 3 &5&2& $(u-3)(u-4)$ & 4,4,-2 & [18,6,9] & [24,12,4] \\
		&   &   &&& $(u-8)$ &   -2,4,4 & LCD& isodual \\
		&   &   &&&  &   4,-2,4 &&  \\\hline

	\end{tabular}}

	\end{document}